# Quantum Dots as Solid-State Sources of Entangled Photon Pairs

Xingling Pan,[1,2] Zhiming Chen,[1,*] Yingtao Ding,[1] Weibo Gao,[6,7] Fei Ding[4,5,*] and Zhaogang Dong[2,3,*]

[1]School of Integrated Circuits and Electronics, Beijing Institute of Technology, Beijing, 100081, P. R. China

[2]Science, Mathematics, and Technology (SMT), Singapore University of Technology and Design, 8 Somapah Road, 487372, Singapore

[3]Quantum Innovation Centre (Q.InC), Agency for Science Technology and Research (A*STAR), 2 Fusionopolis Way, Innovis #08-03, Singapore 138634, Republic of Singapore

[4]School of Electronic Science and Technology, Eastern Institute of Technology, Ningbo, 315200, P. R. China

[5]Centre for Nano Optics, University of Southern Denmark, DK-5230 Odense M, Denmark

[6]School of Electrical and Electronic Engineering, Nanyang Technological University, Singapore 639798, Singapore

[7]Division of Physics and Applied Physics, School of Physical and Mathematical Sciences, Nanyang Technological University, Singapore 639798, Singapore

[*]Authors to whom correspondence should be addressed: zhaogang_dong@sutd.edu.sg; fding@eitech.edu.cn and czm@bit.edu.cn

**Abstract**

Semiconductor quantum dots (QDs) have emerged as a premier solid-state platform for the deterministic generation of nonclassical light, offering a compelling pathway toward scalable quantum photonic systems. While single-photon emission from QDs has reached a high level of maturity, the realization of high-fidelity entangled photon-pair sources remains an active and rapidly evolving frontier. In this review, we survey the recent progress in QD-based entangled photon sources, highlighting the conceptual evolution from the established biexciton–exciton cascade to the emerging paradigm of spontaneous two-photon emission. We further examine how advances in nanophotonic architectures and coherent control strategies are redefining fundamental performance limits, enabling concurrent improvements in source brightness, coherence, and entanglement fidelity. Finally, we discuss the key physical and technological challenges that must be addressed to bridge the gap between laboratory demonstrations and large-scale deployment. We conclude by outlining the future opportunities for integrating QD-based entangled photon sources into practical quantum communication, computation, and sensing platforms.

## I. Introduction

Reliable and bright sources of quantum light constitute a foundational requirement for the development of quantum information technologies. While single-photon sources are rapidly maturing,[1-3] entangled photon-pair sources are critical for applications in quantum communication, photonic quantum computing, and quantum sensing.[4-6] As schematically illustrated in Fig. 1, two principal technological routes have been established for entangled photon-pair generation: spontaneous parametric down-conversion (SPDC) in nonlinear optical media and semiconductor quantum dots (QDs) functioning as solid-state quantum emitters. These two approaches rely on fundamentally different physical mechanisms and therefore exhibit distinct performance trade-offs and target application scenarios.

SPDC exploits energy and momentum conservation in nonlinear crystals and represents a mature and widely adopted platform. It offers high and stable entanglement fidelity, broad spectral tunability, and room-temperature operation, and has been extensively implemented across diverse photonic platforms, including bulk and cavity-enhanced crystals,[7] integrated microresonators,[8] metasurfaces,[5, 9] and emerging low-dimensional and reconfigurable materials.[10-15] These attributes make SPDC particularly suitable for flexible and broadband implementations, ranging from proof-of-principle quantum optics experiments to long-distance free-space and fiber-based quantum communication.[16, 17] However, the intrinsically probabilistic nature of SPDC unavoidably links increased brightness to a higher probability of multi-pair emission, which degrades photon indistinguishability and imposes intrinsic limitations on determinism and scalability, particularly in multi-photon and large-scale quantum photonic architectures.

In contrast, semiconductor QDs provide a solid-state and near-deterministic route to entangled photon-pair generation enabled by their discrete energy-level structures. As shown in Fig. 1(b), QD-based entangled photon emission can proceed via two distinct pathways. In the conventional biexciton–exciton (XX–X) radiative cascade, entangled photon pairs are generated through two sequential first-order transitions, involving real intermediate exciton states. As a result, polarization entanglement is inherently exposed to "which-path" information, rendering the emitted state highly sensitive to excitonic fine-structure splitting (FSS) and solid-state dephasing processes. Alternatively, spontaneous two-photon emission (STPE) constitutes a second-order nonlinear process in which the biexciton decays directly to the ground state via a virtual intermediate state, emitting a photon pair simultaneously.[18] This mechanism intrinsically suppress multi-pair emission and enables high brightness per excitation pulse. Moreover, the compatibility of QDs with nanophotonic platforms supports compact, chip-scale, and potentially scalable implementations. At present, however, large-scale scalability is constrained by challenges associated with deterministic integration and emitter reproducibility. Meanwhile, the indistinguishability and entanglement fidelity of photons emitted from QDs are strongly affected by FSS, solid-state decoherence, and operating conditions, often necessitating careful device engineering and cryogenic operation to reach high entanglement fidelity.[19, 20]

Consequently, SPDC and QDs address complementary application regimes, with SPDC excelling in broadband and flexible implementations, and QDs offering a promising pathway toward deterministic and integrated quantum photonic systems. In this context, a clear understanding of the underlying physical mechanisms and recent experimental progress in QD-based entangled photon sources becomes particularly important. This mini-review therefore surveys recent progress in QD-based entangled photon sources, with a particular focus on the XX–X cascade and the emerging STPE mechanisms, as well as the key challenges and opportunities towards their deployments in practical and scalable quantum photonic platforms.

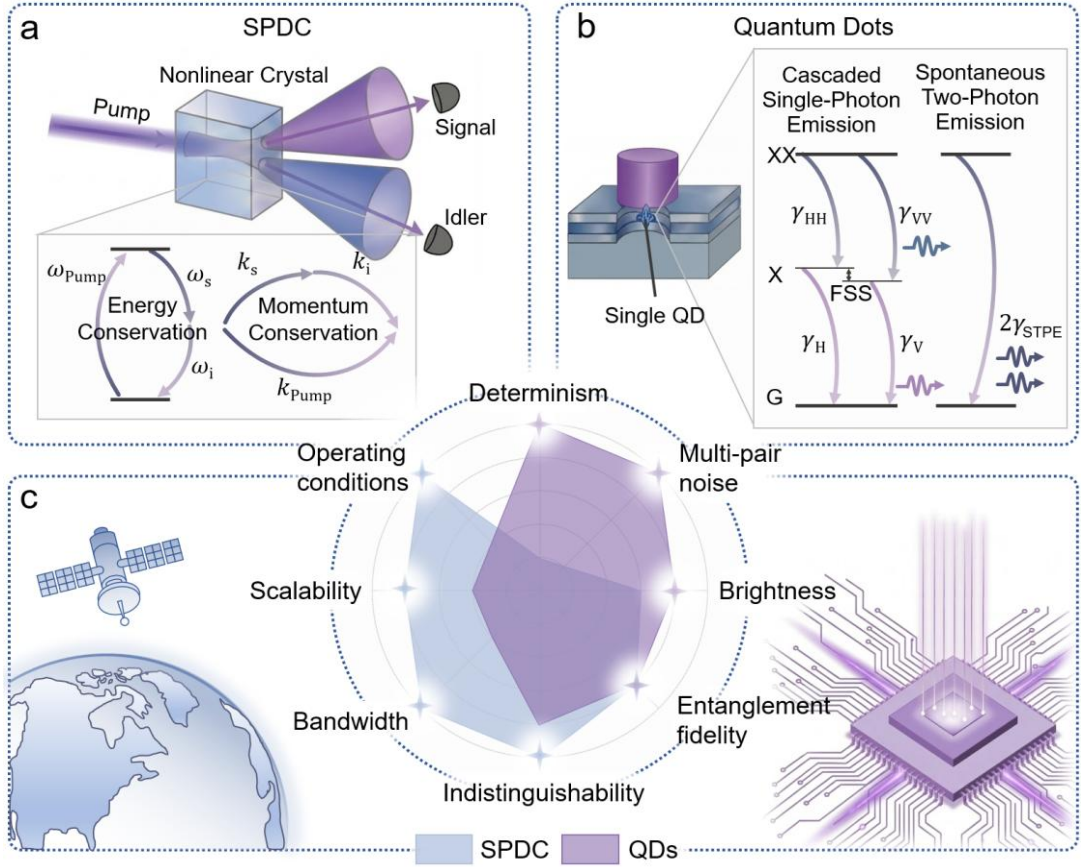

**Figure 1. Comparison of entangled photon pair sources: SPDC vs. semiconductor QDs.** (a) SPDC, a second-order nonlinear optical process in which a pump photon probabilistically splits into a pair of signal and idler photons in a nonlinear crystal, subject to simultaneous energy conservation and phase-matching (momentum conservation) conditions. (b) Semiconductor QD-based entangled photon sources, illustrating a single QD embedded in a nanophotonic structure with its discrete energy-level scheme. Entangled photon pairs are generated either via the biexciton–exciton (XX–X) cascade or spontaneous two-photon emission (STPE). (c) Qualitative performance comparison between SPDC (blue) and quantum dots (purple) across key technology–performance metrics relevant to practical quantum photonic implementations. The radar plot presents a qualitative low–to–high assessment of determinism, multi-pair noise, brightness, entanglement fidelity, photon indistinguishability, spectral bandwidth, scalability, and operating conditions, illustrating how trade-offs among these parameters define the complementary application regimes of the two platforms.

## II. Biexciton–Exciton Cascade Emission in Quantum Dots

The XX–X cascade represents the most established mechanism for generating entangled photon pairs from semiconductor QDs and has long served as the primary experimental benchmark for solid-state entangled photon sources. In this process, a biexciton decays radiatively via an intermediate exciton state, emitting two photons whose polarization correlations can form a Bell state provided that the excitonic FSS is sufficiently small.[21]

Significant efforts have therefore been devoted to improving the performance of XX–X cascade sources through photonic environment engineering. As illustrated in Fig. 2(a), integration of a single QD with a circular Bragg resonator (CBR) enables simultaneous enhancement of photon extraction efficiency and preservation of polarization entanglement. For instance, polarization-resolved X–XX cross-correlation measurements under resonant π-pulse excitation directly confirm high-fidelity entanglement in the linear, diagonal, and circular bases, demonstrating that carefully engineered resonator geometries can support system-level performance metrics combining brightness, indistinguishability, and entanglement fidelity.[22] Moreover, photonic crystal (PC) cavities provide a complementary cavity quantum electrodynamics (CQED)-based route, as shown in Fig. 2(b). Here, strong coupling between excitonic transitions and a high-Q cavity mode leads to pronounced spectral confinement and enhanced light–matter interaction. Beyond improving photon extraction, PC cavities enable access to coherent nonlinear phenomena, such as two-photon Rabi splitting, highlighting the potential of XX–X systems for multi-photon generation and quantum-network-oriented architectures.[23]

In contrast to narrowband cavity-based approaches, broadband photonic interfaces offer an alternative strategy that relaxes stringent resonance-matching conditions. As depicted in Fig. 2(c), dielectric antenna concepts, such as semi-spherical lenses integrated above QDs, enable highly efficient collection of both exciton and biexciton photons over a wide spectral range. Full quantum state tomography confirms that high-quality polarization entanglement can be preserved, emphasizing that broadband mode engineering and detection strategies can compensate for limited spectral selectivity while maintaining strong quantum correlations.[24] Beyond planar cavity geometries, vertically oriented nanowire waveguides offer efficient single-mode emission from embedded quantum dots, naturally combining deterministic positioning, high collection efficiency, and fiber-compatible output.[25-27]

Despite these advances, the XX–X cascade remains fundamentally constrained. Its intrinsic time ordering introduces temporal distinguishability that cannot be fully eliminated, rendering the entanglement fidelity highly sensitive to residual FSS. Intrinsic approaches to suppress FSS, such as employing highly symmetric nanostructures or specific crystal orientations, require stringent fabrication control and are restricted to a narrow set of material systems. To overcome these limitations, post-growth external control technologies have evolved toward providing independent tuning of both the emission energy and FSS.[28-31] A representative example is in situ three-dimensional strain engineering, illustrated in Fig. 2(d).[32] In this approach, an out-of-plane strain component primarily shifts the bandgap and exciton transition energy, while in-plane uniaxial/shear components restore in-plane symmetry to drive a pronounced V-shaped FSS tuning, enabling near-vanishing FSS (≈2 μeV) together with multi-meV wavelength tuning (≈7 meV in CBRs and up to ≈15 meV in micropillars). Complementarily, a dual-Stark scheme [see Fig. 2(e)] combines the quantum-confined Stark effect for emission wavelength tuning via a DC electric field with an optically induced AC Stark shift to selectively modify one exciton eigenstate, thereby cancelling the residual splitting and maintaining high-fidelity entanglement ($f > 0.955$) over ~1 meV wavelength tuning range.[33] While these multi-parameter tuning approaches offer determinism and broad applicability, they inevitably increase device and control complexity, raising practical challenges for scalability, integration density, and long-term robustness.

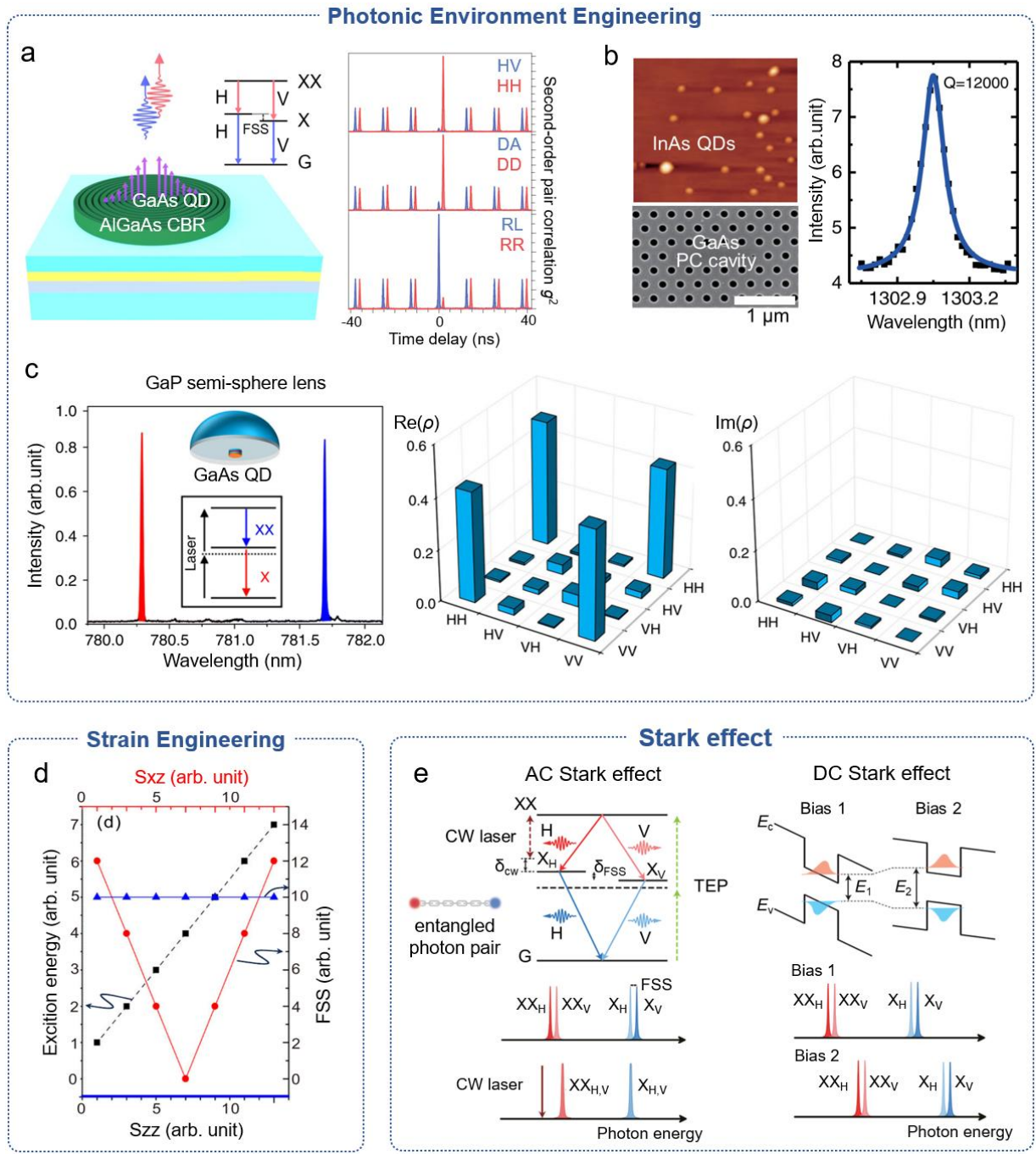

**Figure 2. Biexciton-exciton (XX-X) radiative cascade emission from QDs.** (a) Left panel: Illustration of a circular Bragg resonator (CBR) with a single QD emitting entangled photon pairs. Right panel: Polarization-resolved X–XX cross-correlation histograms measured under π-pulse excitation in the linear, diagonal, and circular bases. Cross-polarized configurations are vertically offset for clarity. The polarization entanglement fidelity of the emitted photon pairs is $f = 0.88 \pm 0.02$. [22] Reproduced with permission from Liu *et al.*, *Nat. Nanotechnol.* 14, 586–593 (2019). Copyright 2019 Springer Nature. (b) Left panel: Atomic force microscope image of QDs within a 1 μm² (top) and scanning electron microscope image of an L3 PC cavity (bottom). Right panel: Representative cavity mode measured at room temperature with Q =

12,000.[23] Reproduced with permission from Qian *et al.*, *Phys. Rev. Lett.* 120, 213901 (2018). Copyright 2018 American Physical Society. (c) Left panel: Highly-efficient extraction of entangled photons from QDs using a broadband optical antenna. Right panel: Real and imaginary parts of the two-photon density matrix obtained from 16 correlation measurements employing the maximum likelihood technique. The fidelity extracted from this matrix is $f = 0.89$.[24] Reproduced with permission from Chen *et al.*, *Nat. Commun.* 9, 2994 (2018). Copyright 2018 Author(s), licensed under a Creative Commons Attribution 4.0 License. (d) The exciton energy and FSS as a function of strain. The exciton energy changes linearly with $S_{zz}$ stress and $S_{xz}$ (black curve). The FSS features a 'V' shaped tuning behavior with $S_{xz}$ stress (red curve), while remaining almost constant with $S_{zz}$ (blue curve).[32] Reproduced with permission from Chen *et al.*, *Nat. Commun.* 16, 5564 (2025). Copyright 2025 Author(s), licensed under a Creative Commons Attribution 4.0 License. (e) Schematic illustration of the combined AC and DC Stark tuning in a single QD. Under AC Stark tuning, a continuous-wave laser induces a polarization-selective energy shift of the exciton states, enabling fine control of the FSS. Under DC Stark tuning, an applied electric field modifies the band bending and shifts the exciton and biexciton transition energies.[33] Reproduced with permission from Chen *et al.*, *Nat. Commun.* 15, 5792 (2024). Copyright 2024 Author(s), licensed under a Creative Commons Attribution 4.0 License.

## III. Emerging Spontaneous Two-Photon Emission Schemes

The limitations of cascaded emission have stimulated growing interest in STPE, which constitutes a more fundamental two-photon emission process. As illustrated in Figure 3(a), STPE involves the direct decay of an initial high-energy state (e.g., the biexciton) into the ground state through the simultaneous emission of a photon pair, bypassing any real intermediate state. This intrinsically quantum process, driven by vacuum fluctuations, offers a potential route to on-demand generation of high-fidelity entangled photons.[34]

Early theoretical studies established that coupling a QD to an optical cavity can strongly enhance the otherwise weak STPE process when the cavity resonance is tuned close to half the biexciton transition energy.[18, 35] This concept was subsequently formalized within master-equation framework describing QD–cavity systems that explicitly incorporated realistic loss and decoherence mechanisms, including radiative decay, pure dephasing, and phonon-assisted processes.[36] Within this unified theoretical framework, the roles of cavity coupling strength, detuning, temperature, and biexciton binding energy in determining the efficiency and entanglement robustness of STPE were systematically clarified,[37] thereby establishing cavity-enhanced STPE as a viable solid-state source of high-quality entangled photon pairs.

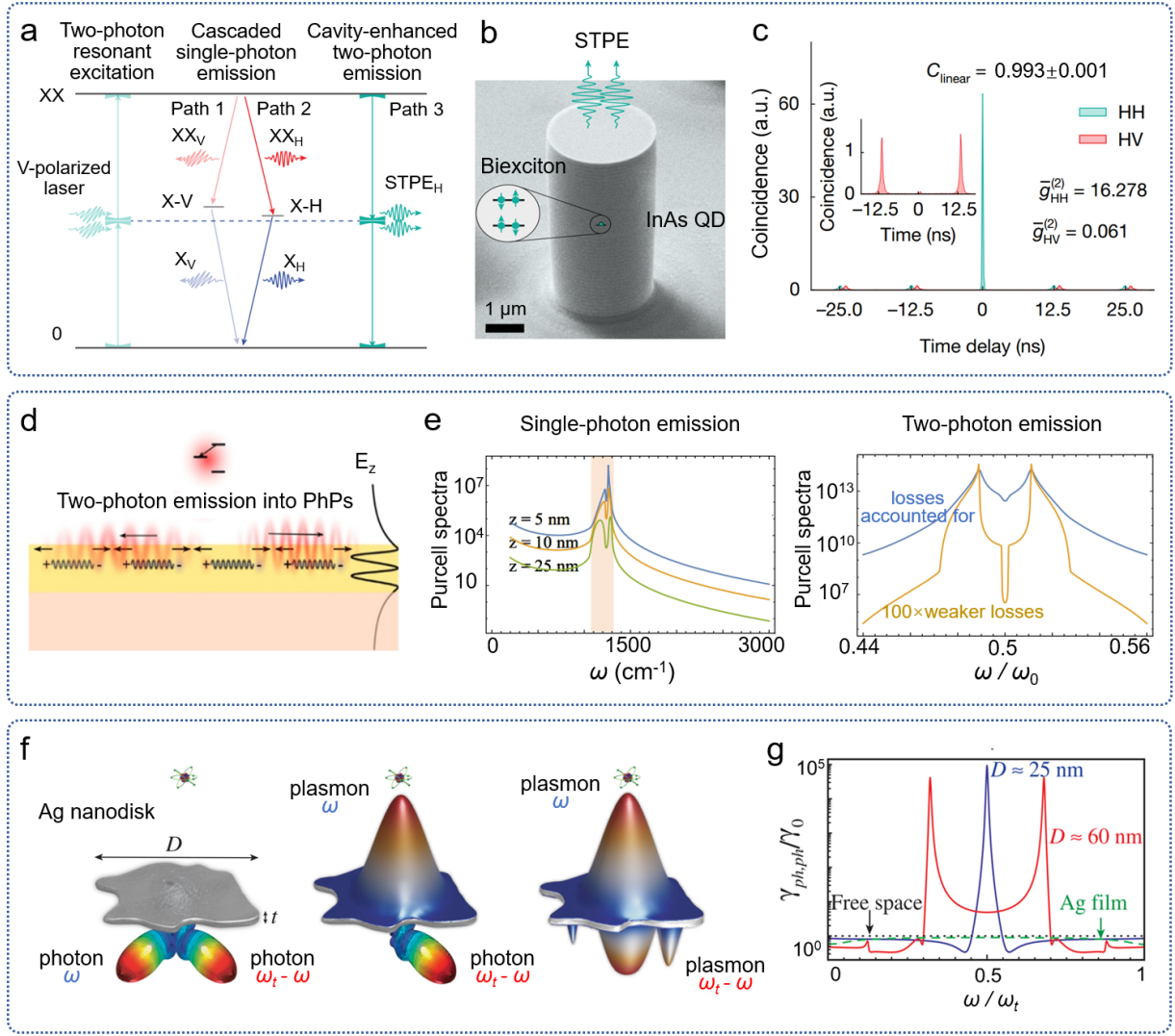

**Figure 3. Emerging spontaneous two-photon emission (STPE) schemes.** (a) Energy-level diagram of STPE in a semiconductor CQED system, illustrating a four-level ladder configuration.[34] (b) SEM image of a micropillar cavity deterministically incorporating a single QD at its center.[34] (c) Polarization-resolved correlation histograms of cavity-enhanced STPE photon pairs measured in the linear polarization basis. Low-probability coincidence events are magnified in the insets for clarify. For compactness, only the linear-basis correlation histogram is shown; the corresponding diagonal and circular bases (not shown) yield consistent near-unity correlations, resulting in an entanglement fidelity of $f = 0.994 \pm 0.001$.[34] Reproduced with permission from Liu *et al.*, *Nature* 643, 1234–1239 (2025). Copyright 2025 Springer Nature. (d) Engineering two-polariton spontaneous emission via phonon polaritons.[38] (e) Single-photon Purcell spectra for a *z*-polarized dipole positioned above a 10-nm-thick cBN film and two-photon Purcell spectra versus photon frequency $\omega$, comparing realistic material losses (blue)

with a reduced-loss scenario (orange). The comparison illustrates phonon-polaritonic enhancement of two-photon spontaneous emission and highlights the critical role of material losses in determining its observable strength.[38] Reproduced with permission from Rivera *et al.*, *Proc. Natl. Acad. Sci. U.S.A.* 114, 13607–13612 (2017). Copyright 2017 Author(s), licensed under a Creative Commons Attribution-NonCommercial-NoDerivatives 4.0 International License. (f) Schematic illustration of the STPE pathways for a quantum emitter close to a two-dimensional plasmonic nanostructure: emission of a photon pair into the far field (left), generation of a hybrid photon-plasmon state (center), or excitation of two plasmonic excitations on the nanostructure (right).[39] (g) Photon-pair production rates for an Ag nanodisk (solid blue and red), an Ag film (green), and free space (black), revealing strong enhancement of photon–photon emission in confined plasmonic geometries.[39] Reproduced with permission from Muniz *et al.*, *Phys. Rev. Lett.* 125, 033601 (2020). Copyright 2020 American Physical Society.

Despite this mature theoretical foundation, experimental realization of STPE remained highly challenging. As an intrinsically second-order radiative process, STPE is strongly suppressed relative to competing single-photon decay channels, such that early experimental observations in individual QDs were limited to weak spectroscopic signatures without direct access to quantum correlations.[35, 40] A landmark achievement was reported by Liu et al. in 2025[34]. In this work, a single InAs QD was deterministically coupled to a high-Q micropillar cavity, with the cavity mode tuned to the two-photon resonance condition, as illustrated in Figure 3(b). Under pulsed two-photon resonant excitation, the detected STPE count rate reached ~497 kHz, which is comparable to the competing single-photon emission rates from the same device via the XX and X channels (~763 kHz and ~813 kHz, respectively). They measured exceptionally high normalized second-order autocorrelation values for $g^2(0)$ (22,691 and 40.26 under CW and pulsed excitations), providing direct evidence of STPE into the same cavity mode. The underlying nonclassical correlations are directly visualized in the polarization-resolved correlation histograms shown in Fig. 3(c). Most importantly, by using a cascaded resonant excitation scheme, polarization-entangled photon pairs were generated with a record-high fidelity of 0.994 ± 0.001. Notably, this fidelity surpasses values reported for the XX–X cascade–based approaches, even when the FSS is effectively eliminated through advanced tuning strategies. The resulting entanglement is intrinsically immune to FSS, thereby combining the near-ideal fidelity characteristic of SPDC sources with the deterministic, on-demand nature of QDs.

Complementary to cavity-based schemes, alternative platforms have been explored to enhance and tailor STPE by engineering light–matter interactions. In photonic waveguides, quadratic coupling between a QD and guided modes has been proposed to enable STPE-like processes, allowing direct generation of frequency-entangled photon pairs through dispersion and mode-density engineering.[41] As shown in Fig. 3(d), polaritonic platforms based on materials, such as hexagonal boron nitride (hBN) and silicon carbide (SiC), provide another route. In these systems, strongly confined phonon polaritons enable an excited emitter to decay via the emission of a correlated polariton pair, allowing two-photon processes to dominate in the mid-infrared regime. Moreover, the corresponding single- and two-photon Purcell spectra in Fig. 3(e) show that, while the single-photon channel mainly serves as a reference, the two-photon Purcell enhancement calculated for an emitter placed at $z$=5 nm exceeds that of single-photon emission by several orders of magnitude. Comparison between realistic and reduced-loss scenarios further demonstrates that material losses critically limit the observable enhancement of the two-photon channel, underscoring dissipation as a decisive design constraint in polaritonic STPE schemes.[38]

In parallel, plasmonic nanostructures offer a complementary pathway by leveraging extreme field confinement and broadband mode densities.[39, 42] For an emitter placed near a finite, two-dimensional plasmonic nanostructure, STPE can proceed through multiple channels, including direct photon–photon emission, hybrid photon–plasmon emission, or plasmon–plasmon generation, as illustrated in Fig. 3(f). In particular, confined plasmonic geometries, such as Ag nanodisks, are able to strongly enhance photon-pair production compared with extended metallic films, highlighting the role of radiative, localized bright plasmon modes in promoting the two-photon channel, as shown in Fig. 3(g). Beyond the platforms discussed above, lithographically scalable resonant nanophotonic architectures, including dielectric Mie resonance, bound-states-in-the-continuum, whispering-gallery, and Fano resonances, are being actively explored to engineer resonant light–matter interactions, offering a potentially powerful route to amplify intrinsically weak two-photon emission channels.[43-46] Beyond platform engineering, dynamic control strategies offer additional flexibility in tailoring photon statistics and correlations associated with two-photon emission. Techniques, such as spectral filtering[47] and coherent field interference,[48] enable the controlled modification of emission properties, further extending STPE from a fundamental quantum-optical process toward a versatile resource for quantum photonic technologies.

## IV. Conclusion and Outlook

QD-based quantum light sources are rapidly evolving from single-photon emitters toward platforms capable of generating increasingly complex multiphoton quantum states. While the XX-X cascade has enabled deterministic entangled photon generation, the recent experimental realization of efficient and high-fidelity entangled photon pair generation via STPE represents a clear conceptual and technological paradigm shift. As summarized in Fig. 4, recent advances have established a diverse set of photonic platforms (see left panel) for tailoring light–matter interactions with QDs, complemented by increasingly sophisticated control strategies, such as field-induced tuning and spectral purification. Together, these developments delineate both emerging opportunities across quantum photonic applications and the key technological challenges that need to be addressed to enable scalable and practical quantum photonic systems.

A central challenge emerging from this landscape is the realization of wafer-scale **deterministic integration** for quantum photonic integrated circuits (QPICs). The CMOS-compatible fabrication of QDs makes them attractive for chip-scale QPICs,[49] while recent advances in pick-place techniques and transfer printing enable heterogeneous integration of III–V QDs onto silicon and silicon nitride photonic platforms.[50] Nevertheless, large scale QPIC implementations critically rely on precise control over QD growth, spatial positioning, and emission uniformity across the wafer. Continued progress in site-controlled epitaxy, deterministic positioning, and system-level co-design of emitters and photonic circuits will therefore be essential for scalable and reproducible QPIC platforms.

Another important direction is the reduction of the strong reliance on **cryogenic operation**. At present, high-fidelity entangled photon emission from QDs typically requires temperatures well below liquid nitrogen, which imposes significant constraints on practical deployment beyond laboratory environments. Future efforts will therefore focus on engineering materials, emission pathways, and photonic environments that are more resilient to phonon induced decoherence, enabling operation even at room temperatures while preserving quantum coherence. In this context, representative approaches include shifting emission control away from coherence-sensitive excitonic processes toward plasmon-mediated or dissipation-engineered mechanisms.[51-53] More broadly, practical operation can also be achieved by moving beyond excitonic coherence altogether and exploiting robust defect spin states with spin-dependent optical readout.[54] Consistently, scalable plasmonic lattices have been employed to enhance room-temperature defect emission in layered materials, such as Al nanotrench arrays that boost the 436-nm hBN B-center emission, underscoring the broad utility of nanophotonic environment engineering under practical conditions.[55]

Enhancing long-term **entanglement stability**, encompassing both quantum-state coherence and device-level reproducibility, is essential for quantum communication networks and quantum information processing. State-of-the-art QD platforms can achieve near unity entanglement fidelity without temporal or spectral post-selection, meeting the requirements of error sensitive protocols, such as quantum key distribution,[56] quantum teleportation,[57] and enabling high rate entanglement distribution in telecom compatible fiber networks.[31, 58] However, practical deployment requires this performance to be maintained in a stable and reproducible manner over extended timescales. Ensuring temporal robustness and device level reproducibility of entangled photon generation therefore remains a central challenge. In parallel, quantum sensing applications exploit source-established nonclassical correlations through coincidence-based measurements and imposes different constraints on quantum-state-level entanglement preservation, relying on reproducible correlation statistics generated at the source.[59] Achieving such stability is likely to require a system-level approach that integrates robust photonic environment engineering with the suppression of slow spectral drift and charge noise, complemented by active stabilization strategies to mitigate environmental fluctuations.

Extending entanglement toward **high-dimensional Hilbert spaces** remains largely unexplored, as most demonstrated entangled photon sources rely on polarization encoding and are therefore restricted to two-dimensional quantum states. Promising routes include exploiting additional photonic degrees-of-freedom, such as orbital angular momentum,[60] time and frequency bins,[61, 62] as well as their combinations through hyperentanglement. In this context, emerging AI-assisted inverse design and optimization techniques offer powerful tools to efficiently engineer nanophotonic structures capable of supporting and controlling increasingly complex, high-dimensional photonic modes.[63, 64] At the same time, metasurfaces provide a powerful physical interface for implementing such control, enabling compact, reconfigurable manipulation of phase, polarization, and mode structure within a single planar platform.[65-69] These approaches can significantly increase information capacity and enhance robustness against noise and loss, while remaining compatible with fiber-based transmission and integrated photonic platforms. Beyond communication, high-dimensional encodings can also benefit photonic quantum computing architectures based on discrete-variable (DV) photons, where qudit encodings enable more efficient use of photonic degrees-of-freedom and can reduce circuit complexity and resource overhead.[70-72] This progression motivates the development of hybrid discrete-variable–continuous-variable (DV–CV) photonic architectures, in which discrete photonic qudits are interfaced with bosonic modes for encoding or error-mitigation purposes,[73, 74] further motivating the integration of deterministic quantum emitters with advanced mode-control platforms.[75]

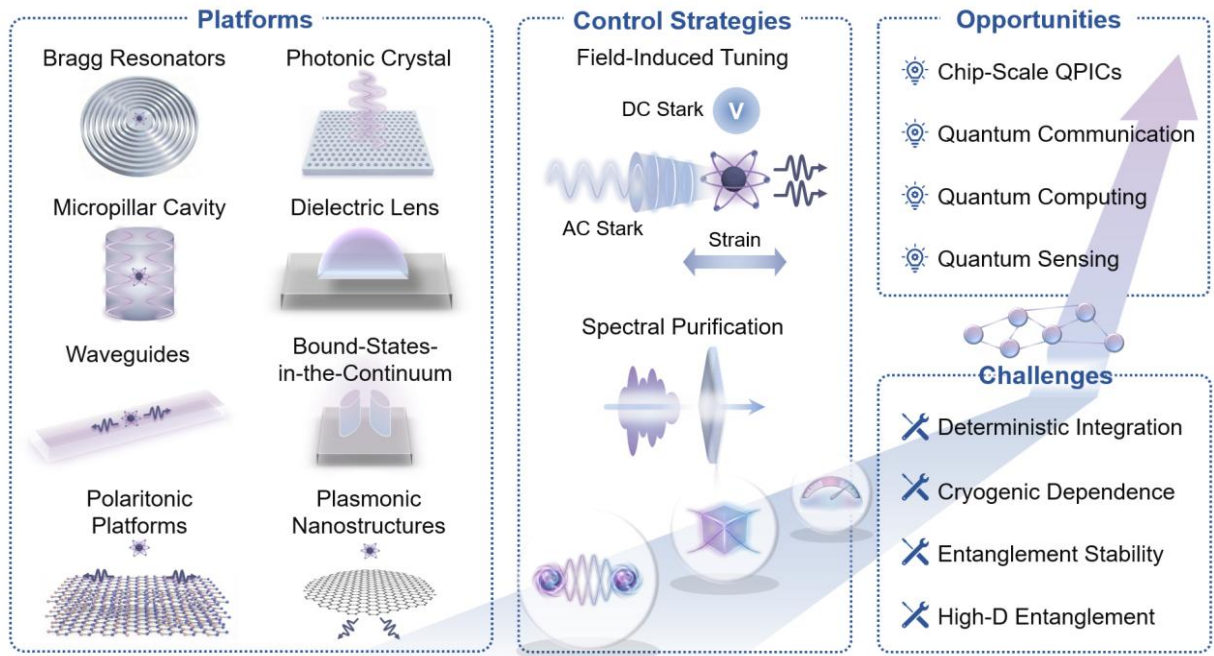

**Figure 4. Outlook on QDs as solid-state sources of entangled photon pairs.** Left panel: a broad range of nanophotonic environments is presented, illustrating how tailored light–matter interactions enhance emission efficiency, mode control, and photon extraction. Middle panel: Advanced control strategies provide deterministic control over FSS, emission energy, and photon statistics, enabling both XX–X cascade emission and emerging STPE schemes. Right panel: Application opportunities in chip-scale quantum photonic integrated circuits (QPICs), quantum communication, quantum computing, and quantum sensing are highlighted, together with the remaining challenges including wafer-scale deterministic integration, cryogenic operation, long-term entanglement stability, and extension of entanglement toward high-dimensional Hilbert spaces.

Overall, QD-based entangle photon sources are transitioning from proof-of-concept demonstrations toward scalable and application-oriented quantum photonic technologies. The convergence of new emission mechanisms, advanced nanophotonic platforms, and increasingly sophisticated control strategies is steadily addressing long-standing limitations in efficiency, fidelity, and functionality. Meanwhile, realizing the full potential of these sources will require coordinated progress in wafer-scale deterministic integration, operation at technologically accessible temperatures, long-term entanglement stability, and access to high-dimensional quantum states. Continued advances along these directions are expected to establish QDs as robust and versatile building blocks for QPICs and future quantum communication, computation, and sensing systems.

**Acknowledgments**

Z. D. would like to acknowledge the funding support from The National Research Foundation (NRF), Singapore via Grant No. NRF-CRP30-2023-0003, and the Agency for Science, Technology and Research (A*STAR) under its MTC IRG (Project No. M24N7c0083) and its Japan–Singapore Joint Call for Quantum 2025 (Grant No. R25J4IR112). F. D. would like to acknowledge the Zhejiang Provincial Natural Science Foundation of China (LR26F050003), and Carlsberg Foundation (CF24-1777). Z. C. and Y. D. would like to acknowledge the National Natural Science Foundation of China (92373105 and 62074015). X. P. would like to acknowledge the support from the China Scholarship Council (Grant No. 202406030248).